\documentclass[11pt]{article}
\usepackage{epsfig}
\newcommand{\la}[1]{\label{#1}}
\newcommand{\be}{\begin{equation}}
\newcommand{\ee}{\end{equation}}
\newcommand{\ba}{\begin{eqnarray}}
\newcommand{\ea}{\end{eqnarray}}
\newcommand{\rmi}[1]{{\mbox{\scriptsize #1}}}
\newcommand{\fig}{Fig.~}
\newcommand{\eq}{Eq.~}
\newcommand{\se}{Sec.~}
\newcommand{\eqs}{Eqs.~}
\newcommand{\nr}[1]{(\ref{#1})}

\newcommand{\fr}[2]{{\frac{#1}{#2}\,}}

\renewcommand{\vec}[1]{{\bf #1}}

\def\lsi{\raise0.3ex\hbox{$<$\kern-0.75em\raise-1.1ex\hbox{$\sim$}}}
\def\gsi{\raise0.3ex\hbox{$>$\kern-0.75em\raise-1.1ex\hbox{$\sim$}}}

\makeatletter \@addtoreset{equation}{section} \makeatother

\makeatletter
\renewcommand\section{\@startsection {section}{1}{\z@}%
                                   {-5.5ex \@plus -1ex \@minus -.2ex}
                                   {2.3ex \@plus.2ex}%
                                   {\normalfont\large\bfseries}}
\renewcommand\subsection{\@startsection{subsection}{2}{\z@}%
                                     {-3.25ex\@plus -1ex \@minus -.2ex}%
                                     {1.5ex \@plus .2ex}%
                                     {\normalfont\normalsize\bfseries}}
\renewcommand\thesection {\@arabic\c@section}
\renewcommand\thesubsection   {\thesection.\@arabic\c@subsection}
\renewcommand{\@seccntformat}[1]{%
\csname the#1\endcsname.\hspace{1.0em}}
\makeatother

\begin{document}

\begin{titlepage}

\begin{flushright}
CERN-TH/2003-092\\
HIP-2003-23/TH\\
\end{flushright}

\begin{centering}
\vfill

{\Large{\bf Q-ball dynamics from atomic Bose-Einstein condensates}}

\vspace{0.8cm}

K.~Enqvist$^{\rm a,}$\footnote{kari.enqvist@helsinki.fi} and
M.~Laine$^{\rm b,}$\footnote{mikko.laine@cern.ch}$^{,}$%
\footnote{Address since June 1, 2003: Faculty of Physics, 
University of Bielefeld, D-33501 Bielefeld, Germany.}
\\
\vspace{0.8cm}
{\em $^{\rm a}$%
Dept.\ of Physical Sciences, University of Helsinki,
and Helsinki Institute of Physics, \\
P.O.Box 64, FIN-00014 University of Helsinki, Finland\\}
\vspace{0.3cm}
{\em $^{\rm b}$%
Theory Division, CERN, CH-1211 Geneva 23,
Switzerland\\}
\vspace*{0.8cm}

{\bf Abstract}

\vspace*{0.3cm}

\end{centering}

\noindent
Relativistic scalar field theories with a conserved global charge $Q$
possess often (meta)stable spherically symmetric soliton
solutions, called Q-balls. We elaborate on the perfect formal analogy
which exists between Q-balls, and spherically symmetric solitons in
certain non-relativistic atomic Bose-Einstein condensates, for which
the dominant interatomic interaction can be tuned attractive.
In a harmonic trap, present in existing
experiments, the Q-ball solution is modified in an essential way. If the
trap is significantly prolongated in one direction, however, then genuine
solitons do appear, and actual experimental data can be obtained for
some of the Q-ball properties studied numerically in the 
relativistic cosmological context, such as their formation and
collisions. We also suggest conditions under which the same
cosmologically relevant analogies could be extended to the fully
three-dimensional case. 
\vfill
\noindent

%
%
%
%

\vspace*{1cm}

\noindent
July 2003

\vfill

\end{titlepage}

%
\section{Introduction}
\la{se:introduction}

Solitons, (meta)stable non-dispersive bound states,
appear in all areas of physics described by non-linear field theories.
They can be considered as lumps of spatially localised matter.
As is well-known, solitons can be topological or non-topological in
nature. Here we shall be concerned with non-topological solitons, which
are defined as field configurations having the same boundary conditions
at asymptotic infinity as the vacuum state. For an extensive review on
non-topological solitons in both relativistic and non-relativistic field
theories, see~\cite{lp}.

Most soliton solutions exist in one space-dimension only, as required
by a theorem due to Derrick~\cite{derrick}. The theorem can, however, be
circumvented, either by including gauge fields of non-zero spin or,
as will be discussed here, by considering time-dependent solutions.
The bottom line is that stable spherically symmetric non-topological
solitons~\cite{rosen,fls}, also called Q-balls~\cite{qball},
can exist if the theory contains scalar fields
with suitable self-interactions, a conserved particle
number, or ``charge'' $Q$, and the charge is carried by massive particles.

In particle physics and cosmology,
Q-balls have attracted a lot of attention recently.
The reason is that the requirements for their existence are satisfied by
(approximately) supersymmetric theories~\cite{mssm}, considered among
the best alternatives for physics Beyond the Standard Model.
Indeed, such theories do have new scalar fields, in the form e.g.\
of ``squarks''.
In this case the conserved charge is the baryon number $B$
(or some combination of the baryon and lepton numbers).
The precise properties of Q-balls depend on the particular model
of supersymmetry breaking, but for many conceivable alternatives,
supersymmetry-based Q-balls may contribute significantly
to the dark matter~\cite{ks} and baryon contents~\cite{emc,ls,bj}
of the Universe (for recent reviews, see~\cite{em}).
Stable Q-balls can also be directly searched for in existing
and planned experiments~\cite{kkst}.

In the present paper we transport the Q-ball formalism to
non-relativistic atomic Bose-Einstein Condensates (BECs), which
during the last few years have been the subject of exciting experimental
developments. Our purpose is to elaborate on the formal analogy
that exists between Q-balls and various solitons in BECs
(see~\cite{lp,ss} for reviews).
We thus discuss, on one hand, whether it might be possible
to observe spherically symmetric three-dimensional (3D) Q-balls
in actual BEC experiments and, on the other hand, what kind of
analogies can be drawn from the already existing experiments
with essentially one-dimensional (1D) solitons,
to support numerical studies of their dynamics. Our hope is that
these analogies might allow to obtain some insights
on the behaviour of Q-balls also in cosmology.

%
\section{Q-balls in relativistic field theory}

In order to set up the relation to the non-relativistic case, 
let us start by briefly reiterating the properties of
non-topological solitons, or Q-balls,
at zero temperature in a relativistic field theory
(for a recent review see, e.g.,~\cite{mgs})\footnote{%
  In this section we employ the conventional natural units $\hbar = c = 1$.
  }.
Consider a generic field theory containing a scalar
field $\phi$ and having a global U(1) symmetry.
Let us denote the scalar potential by $U(|\phi|)$.
The Minkowskian action is then
\be
 S_M = \int {\rm d}t \, {\rm d}^3 x
 \Bigl[
 |\partial_t\phi|^2 -
 |\nabla \phi|^2 - U(|\phi|)
 \Bigr]
 \;.
 \la{rSM}
\ee
According to the Noether theorem, the system described by this action
possesses a conserved ``charge'', $Q$,
\be
 Q = \int {\rm d}^3 x
 \, \Bigl[
 i \Bigl(
 \phi^* \partial_t \phi - \phi\,\partial_t \phi^*
 \Bigr)
 \Bigr]
 \;.
 \la{rQ}
\ee
In the non-relativistic case $Q$ could be the number of atoms, while
in the relativistic case it could be the baryon number, as in supersymmetric
theories.

The question is, what kind of solutions are there for the classical
equations of motions derived from $S_M$,
given some fixed value of $Q$? In order
to answer this question it is convenient to introduce a Lagrange
multiplier $\mu$ conjugate to $Q$, and consider the expression for
the energy of the system in the sector of a fixed $\mu$
first\footnote{%
  Oftentimes the notation $\omega$ is used instead of $\mu$, but
  in a thermodynamic sense the quantity in question is precisely
  the chemical potential for $Q$, which is why we prefer this notation.
  } (see, e.g., \cite{lp,smallQ}).
Then, a Q-ball solution has the
form~\cite{rosen,fls}
\be
 \phi(\vec{x},t)=\exp(-i \mu t) \phi(\vec{x}).
\ee
The energy related to this configuration in the given ensemble is
\be
 \Omega(\mu) = \int {\rm d}^3x \,
 \Bigl[ |\nabla\phi|^2
  - \mu^2 |\phi|^2 + U(|\phi|) \Bigr]
 \;. \la{rOM}
\ee
The chemical potential $\mu$ is related to the total charge
of the solution $Q$ through
\be
 Q = -\frac{\partial \Omega(\mu)}{\partial\mu}=
 2\mu\int {\rm d}^3x \, |\phi|^2
 \;.
 \la{rQ2}
\ee
Finally, the energy in the sector of
a fixed charge is obtained by a Legendre transform,
\be
 E(Q) = \Omega(\mu) + \mu Q\;,
 \la{rEQ}
\ee
where $\mu$ is expressed in terms of $Q$ by inverting~\eq\nr{rQ2}.

To study whether $\Omega(\mu)$ has non-trivial extrema,
one writes
\be
 \phi(\vec{x}) = \frac{1}{\sqrt{2}} v(\vec{x}) e^{i \alpha(\vec{x})}
 \;.
 \la{rFORM}
\ee
A spatial variation in
$\alpha(\vec{x})$ costs energy, so that we may assume
it a constant, and without loss of generality, choose it to vanish.
Moreover, energy is minimised by a spherically
symmetric configuration~\cite{cgm}. 
The corresponding profile $v(r)$ is then determined by
the classical equation of motion following from $\Omega(\mu)$,
\be
 \biggl(
 \frac{{\rm d}^2}{{\rm d} r^2} + \frac{2}{r}
 \frac{{\rm d}}{{\rm d}r}
 \biggr) v(r) =
 -\mu^2 v + \frac{{\rm d} U(v/\sqrt{2})}{{\rm d}v}
 \;.
 \la{rEOM}
\ee
By choosing a suitable $\mu$
this system {\it does have} a non-trivial localised solution,
called a Q-ball, provided that
$U({v}/\sqrt{2})/v^2$ has a minimum
for some $v>0$~\cite{rosen,fls,qball}. Physically, this implies
the existence of an attractive interaction which can bind a
collection of elementary quanta into a
condensate, or ``lump'', of semiclassical matter.

The fact that a solution exists for the classical equation of
motion does not yet guarantee that
it is stable with respect to quantum
fluctuations. This is the case only if the energy of
the solution, $E(Q)$, is smaller than the energy of an ensemble
of free particles of mass $m$ carrying the same charge:
\be
 E(Q) < m Q
 \;.
 \la{stability}
\ee
Under these conditions Q-balls are absolutely stable~\cite{fls}.
If \eq\nr{stability} is not satisfied, then Q-balls
can still be metastable~\cite{fls}
but possibly long-lived (see, e.g.,~\cite{evap}).

These basic considerations can be refined in a number of ways.
For instance, finite temperature corrections can be addressed
through the grand canonical potential, $\Omega(T,V,\mu)$.
It is a standard procedure to derive a Euclidean
(``imaginary time'') path integral expression for $\Omega(T,V,\mu)$,
and one can generically carry out also ``dimensional reduction'' in this
expression, integrating out the non-zero Fourier modes for the
dependence of the fields on the time-like coordinate. The result
is just~\eq\nr{rOM}, only with modified parameters, containing
now all relevant dependence on the temperature $T$~\cite{ls}.
One can also address a wide variety of different potentials:
for instance, if the potential is ``flat'' at large $|\phi|$,
modulo possible logarithmic corrections, then the
energy of the solution $E(Q)$ scales as
$E(Q) \sim M Q^{3/4}$~\cite{fls,dks}, allowing to satisfy~\eq\nr{stability}
for $Q \gg (M/m)^4$, and making Q-balls absolutely stable. With
other potentials the growth may be slower than $|\phi|^2$
only by radiatively induced logarithmic corrections~\cite{ejm}, but it
is still possible to find regions in the parameter space where
Q-balls are absolutely stable~\cite{stable}. Finally,
it is possible to address the formation of Q-balls from
the fragmentation~\cite{ks,emc,formation} of an essentially
homogeneous initial condensate~\cite{ad} as well as, in case Q-balls
are only metastable, their decays and lifetime~\cite{evap},
particularly at finite temperatures~\cite{ks,emc,ls,bj}.

%
\section{Q-balls in non-relativistic field theory}

Let us now turn to the non-relativistic case appropriate for atomic BECs.
As is conventional in this context, we denote the scalar field by $\psi$
instead of $\phi$, and reintroduce $\hbar$.

The ``vacuum'' action describing the weakly interacting
atoms can be written as
\be
 S_M = \int {\rm d}t\, {\rm d}^3 x
 \biggl\{
 i\hbar\,  \psi^* \partial_t \psi
 - \frac{\hbar^2}{2 m} |\nabla\psi|^2 -
 \Bigl[
 V_0  + V(\vec{x})
 \Bigr]
 |\psi|^2
 - \frac{2\pi\hbar^2 a}{m} |\psi|^4 - ...
 \biggr\}
 \;.
 \la{nrSM}
\ee
Here $m$ is the atom mass, $a$ is the s-wave scattering length,
and $V(\vec{x})$ is a possible external potential.
The conserved Noether charge corresponding to this theory
is just the particle number,
\be
 Q = \int {\rm d}^3 x\, |\psi|^2 \;,
 \la{nrQ}
\ee
and the energy for stationary configurations, obtained for an ensemble
with a chemical potential $\mu$ conjugate to $Q$, is then
\be
 \Omega(\mu) = \int {\rm d}^3 x \,
 \biggl\{
 \frac{\hbar^2}{2 m} |\nabla \psi|^2 +
 \Bigl[ - \mu + V(\vec{x})\Bigr] |\psi|^2
 + \frac{2\pi\hbar^2 a}{m} |\psi|^4 + ...
 \biggr\} \;.
 \la{nrOM}
\ee
As is conventional, the chemical potential has been additively
redefined here such that it contains the part $V_0\sim m$ in~\eq\nr{nrSM}.
The energy for a fixed charge is again obtained from
\be
 E(Q) = \Omega(\mu) + \mu Q\;,
 \la{nrEQ}
\ee
where $\mu$ is expressed in terms of $Q$ by inverting~\eq\nr{nrQ}
for a given solution $\psi$ depending on $\mu$.
Let us remark that finite temperature effects could be taken
into account in complete analogy with the relativistic case:
one can again write down a Euclidean path integral expression for the
grand canonical potential and carry out dimensional reduction,
to arrive at an expression of precisely the form in~\eq\nr{nrOM},
only with modified parameter values~\cite{at}.

The equation of motion following from
$\Omega(\mu)$ is the (stationary) Gross-Pitaevskii equation~\cite{gp},
\be
 \biggl[
 -\frac{\hbar^2}{2 m}\nabla^2 - \mu + V(\vec{x}) + \frac{4\pi\hbar^2a}{m}
 |\psi|^2 + ...
 \biggr] \psi = 0 \;.
 \la{nrEOM}
\ee
We observe that (apart from trivial changes)
\eqs\nr{nrOM}, \nr{nrEOM} describe precisely the same physics
as~\eqs\nr{rOM}, \nr{rEOM}, if $V(\vec{x})=0$.
Note that in the relativistic case, $U(|\phi|)$
includes also a term quadratic in $\phi$,
$U(|\phi|) = m^2 |\phi|^2 + ...\;$,
so that $m^2 - \mu^2_\rmi{rel} \sim -2\, m \,\mu_\rmi{non-rel}$.

Since~\eqs\nr{rOM}, \nr{nrOM} are equivalent for
the homogeneous case $ V(\vec{x}) = 0$, the conditions
for the existence of (meta)stable Q-ball
solutions are also equivalent: the potential needs to grow more
slowly than $|\psi|^2$. This can
be achieved if there is an attractive interaction between the atoms
or, equivalently, if the s-wave scattering length
is negative, $a < 0$. This is indeed
the case for instance for the alkali vapour $^7$Li~\cite{bsth}. More
generally, the magnitude of $a$ in BECs
can be tuned in a wide range, including
both positive and negative values, using a magnetic field close to
a so-called Feshbach resonance (see, e.g.,~\cite{fb}), 
as has been demonstrated
also for $^{23}$Na~\cite{ias}, $^{85}$Rb~\cite{rb}, and 
$^{133}$Cs~\cite{cs}.
In the following, we thus assume that $a < 0$.

Obviously, setting just $a < 0$ in~\eq\nr{nrOM} is
somewhat discomforting, because the theory is then not
well-defined, being unbounded from below. This implies
that the system tends to undergo a phase transition to the true
ground state, possibly a Bose liquid~\cite{ss,htcs}. It is observed
experimentally, however, that at least on short enough time scales
a weakly interacting gaseous phase is still present, even when $a < 0$.

The theory in \eq\nr{nrOM} can be explicitly stabilized, however,
by adding higher order operators, for instance~\cite{bhh}
\be
 \delta \Omega(\mu) =
 \int {\rm d}^3 x\,\biggl\{
 A |\nabla (\psi^* \psi)|^2 + B (\psi^*\psi)^3
 \biggr\}
 \;,
 \la{higherops}
\ee
where $A$ parameterises the effective range of the two-body scattering
problem, and $B$ the amplitude for three-body collisions.
Relativistic Q-ball solutions in the case that $a$ is negative
but $B$ is non-vanishing, have been discussed in~\cite{bw}.
On the other hand,
there is a range of chemical potentials where we are in the region
of the ``thick-wall approximation'', and any stabilising terms,
such as $B$, can be neglected~\cite{smallQ}. In the following
we will for simplicity ignore $A$ but keep $B$, in order to
understand when effects from operators such as those
in~\eq\nr{higherops} are important.
Note that in the dilute and (almost) homogeneous limit the operator
multiplied by $B$ can be argued to be parametrically more important 
than that multiplied by $A$~\cite{bhh}.

%
\section{Solution in homogeneous space}
\la{se:homo}

Let us now consider in more detail the non-relativistic
but {\em homogeneous} case,
that is $V(\mathbf{x})\equiv 0$ in~\eqs\nr{nrSM}, \nr{nrOM}, \nr{nrEOM},
but $B\neq 0$ in~\eq\nr{higherops}.
The solution resembles very much the relativistic one discussed in~\cite{bw},
the main difference being in the relation of $\mu$ and $Q$, but for 
completeness and since the solution does not appear to be widely 
appreciated in the atomic BEC literature, we briefly present 
some of its main features 
here, using the notation conventional in that context. 

\subsection{Equations of motion}

As in~\eq\nr{rFORM}, we can write the solution
of~\eq\nr{nrEOM} in the form
\be
 \psi = \frac{1}{\sqrt{2}} v e^{i \alpha} \;,
\ee
where $v \ge 0$ and $\alpha$ can be chosen to vanish.
\eq\nr{nrEOM} then takes the form corresponding to~\eq\nr{rEOM},
\be
 \biggl[
 -\frac{\hbar^2}{2m}
 \biggl(
 \frac{{\rm d}^2}{{\rm d} r^2} +
 \frac{2}{r} \frac{{\rm d}}{{\rm d} r}
 \biggr)
 -\mu + \frac{2\pi \hbar^2 a}{ m} v^2 + \fr34 B v^4
 \biggr] v = 0 \;.
 \la{norGP}
\ee
The boundary conditions are that
\ba
 & & v'(0) = 0; \quad
 \lim_{r\to \infty} v(r) = \lim_{r\to \infty} v'(r) = 0 \;.
 \la{norBC}
\ea
Since the system is over-constrained, solutions are only found
for specific values of $v(0)$.

The parametric dependences of all the properties of
the solution of~\eqs\nr{norGP}, \nr{norBC} can easily be found
out. A non-trivial solution exists for $\mu < 0$, and in the attractive case
we are interested in here, $a < 0$. We can then rescale
\be
 r \equiv \hat r\, a_r, \quad
 v \equiv \hat v\, a_v, \quad
 B \equiv \beta \, a_B
 \;,
 \la{rescale}
\ee
with
\be
 a_r \equiv \sqrt{\frac{\hbar^2}{2 m |\mu|}}, \quad
 a_v \equiv \sqrt{\frac{m |\mu|}{2 \pi \hbar^2 |a|}}, \quad
 a_B \equiv \biggl(\frac{\pi\hbar^2|a|}{m}\biggr)^2 \frac{1}{|\mu|}
 \;,
 \la{rescale2}
\ee
whereby a common factor $|\mu|$ can be dropped out from the equation.
After this rescaling, \eq\nr{norGP} becomes
\be
 \biggl(
 \frac{{\rm d}^2}{{\rm d} {\hat r}^2} +
 \frac{2}{\hat r} \frac{{\rm d}}{{\rm d} \hat r}
 \biggr) \hat v = \hat v - {\hat v}^3 + \frac{3}{16} \beta\, {\hat v}^5\;.
 \la{simGP}
\ee
Given the solution with the boundary conditions
corresponding to \eq\nr{norBC}, we can compute
the dimensionless equivalents of \eqs\nr{nrQ}, \nr{nrOM}, \nr{nrEQ}:
\ba
 Q_\beta & \equiv & \beta^{-\fr12} 
 \int_0^\infty {\rm d} \hat r \, 4\pi {\hat r}^2  \,
 \cdot \fr12 {\hat v}^2
 \;, \la{Qbeta} \\
 \Omega_\beta & \equiv & \beta^{\fr12} 
 \int_0^\infty {\rm d} \hat r \, 4\pi {\hat r}^2  \,
 \cdot \biggl[
 \fr12 |\nabla {\hat v}|^2
 + \fr12 {\hat v}^2
 - \fr14 {\hat v}^4
 + \frac{\beta}{32} {\hat v}^6
 \biggr]
 \;, \\
 E_\beta & \equiv &
 \Omega_\beta - \beta\, Q_\beta
 \;. \la{Ebeta}
\ea
Factors of $\beta$ have been chosen such that rescalings back
to physical units contain no $|\mu|$'s, other than implicitly
inside the $\beta$ in $Q_\beta, E_\beta$:
\ba
 Q(|\mu|) & = &
 \Bigl( \frac{m B}{2\hbar^2} \Bigr)^{1/2}
 \Bigl( \frac{1}{2 \pi |a|} \Bigr)^{2}
 Q_\beta \;,
 \la{rsQmu}
 \\
 E(Q) & = &
 \Bigl( \frac{1}{2 m B} \Bigr)^{1/2}
 \frac{\hbar^3}{4 m} E_\beta
 \;.
\ea

\begin{figure}[t]

\begin{center}
\epsfig{file=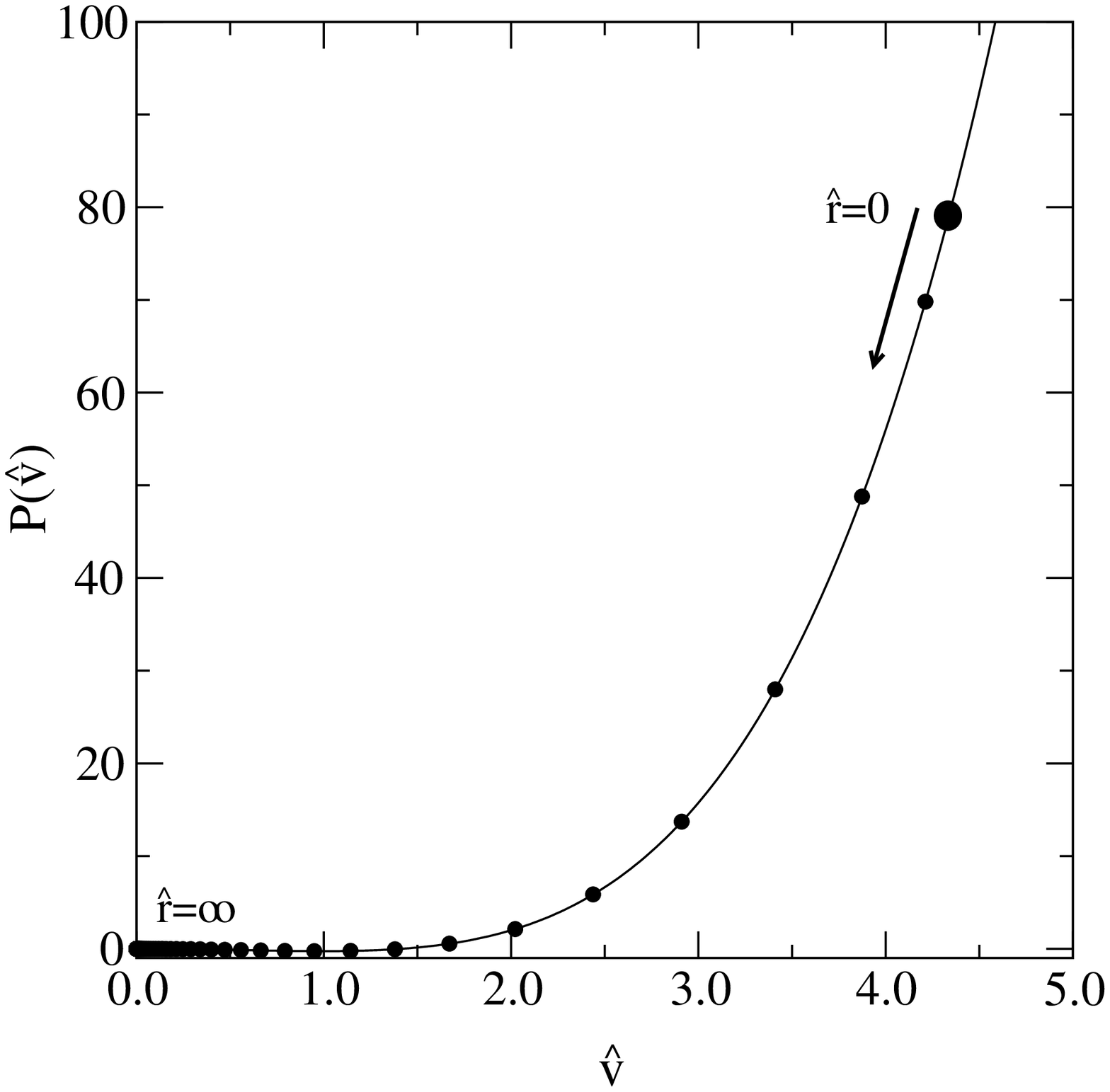,width=7cm}%
\hspace*{1cm}%
\epsfig{file=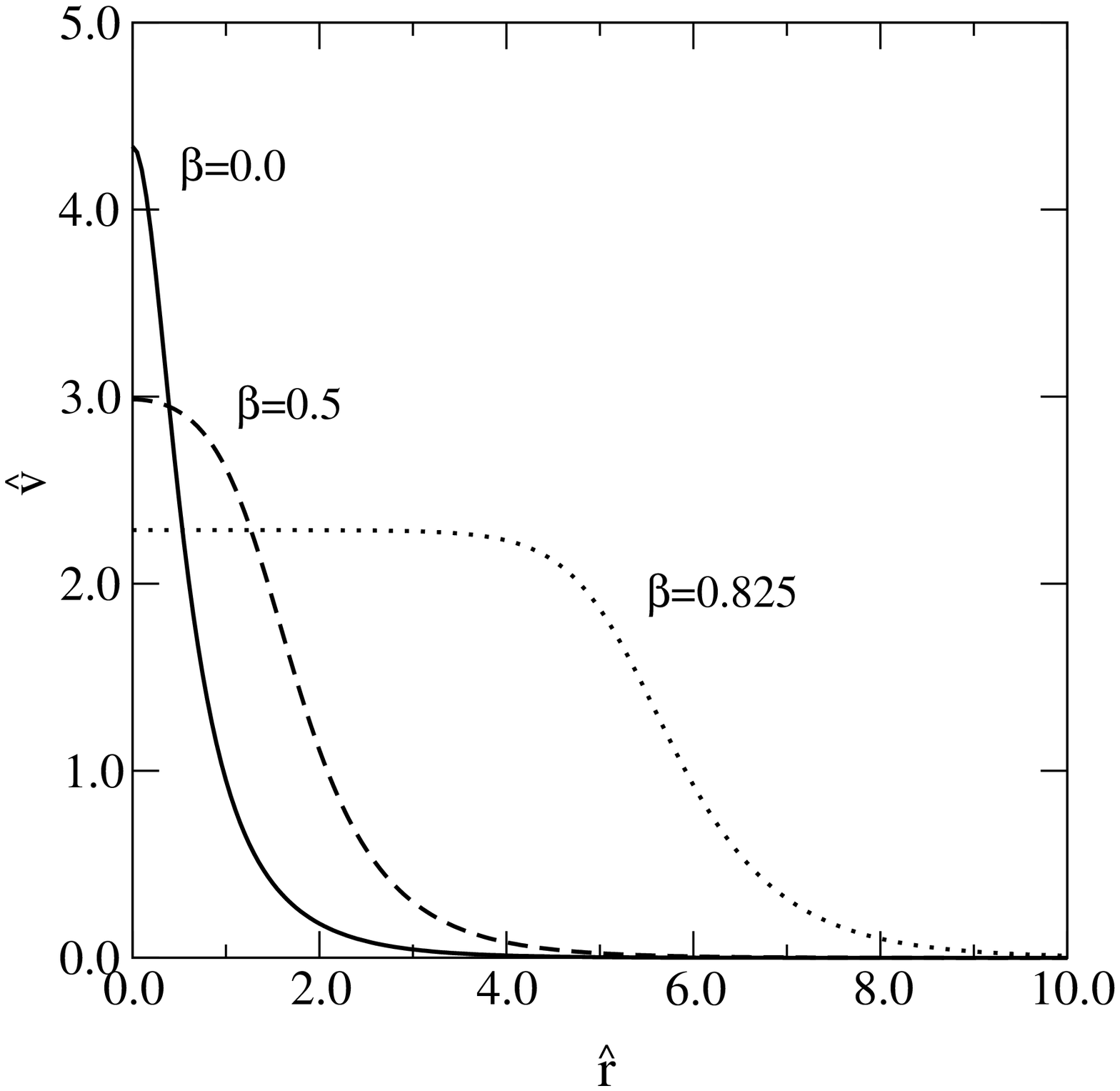,width=7cm}%
\end{center}

\caption[a]{Left: the Q-ball solution for the normalised (inverted)
potential $P(\hat v) = - {\hat v}^2/2 + {\hat v}^4/4$, corresponding
to~\eq\nr{simGP} with $\beta=0$.
The dots are equidistant
in $\hat r$, corresponding to uniform time intervals in the
classical mechanics analogue.
Right: the corresponding profiles $\hat v(\hat r)$,
for $\beta=0.0, 0.5, 0.825$.}

\la{fig:solution}
\end{figure}

\subsection{Exact numerical solution}

It is well-known~(cf.\ ref.~\cite{fls})
that a non-trivial solution exists for~\eq\nr{simGP}
with the boundary conditions of~\eq\nr{norBC},
for $\beta \le 1$. (Because we want the theory to be stable,
we also assume $\beta \ge 0$.)
We may recall that
the simplest way to understand this is to think of $\hat r$
in \eq\nr{simGP} as a time variable, $\hat v$ as a position,
and to note that~\eq\nr{simGP} then corresponds to the movement
of a classical particle in a potential
$P(\hat v) = -\fr12 {\hat v}^2 + \fr14 {\hat v}^4 -
\frac{\beta}{32} {\hat v}^6$,
under the influence of some friction. The situation is illustrated
in~\fig\ref{fig:solution}(left), for $\beta=0$.
A solution can be found by a simple
overshooting-undershooting algorithm, and is also illustrated
in~\fig\ref{fig:solution}.
We find the range
\be
 \hat v ( 0 ) = 4.3374\, ... \, 2.0
 \;,
 \la{numerical}
\ee
for $\beta = 0.0\, ...\, 1.0$, respectively.
The corresponding $Q_\beta, E_\beta$
are shown in~\fig\ref{fig:branches}.

\begin{figure}[t]

\begin{center}
\epsfig{file=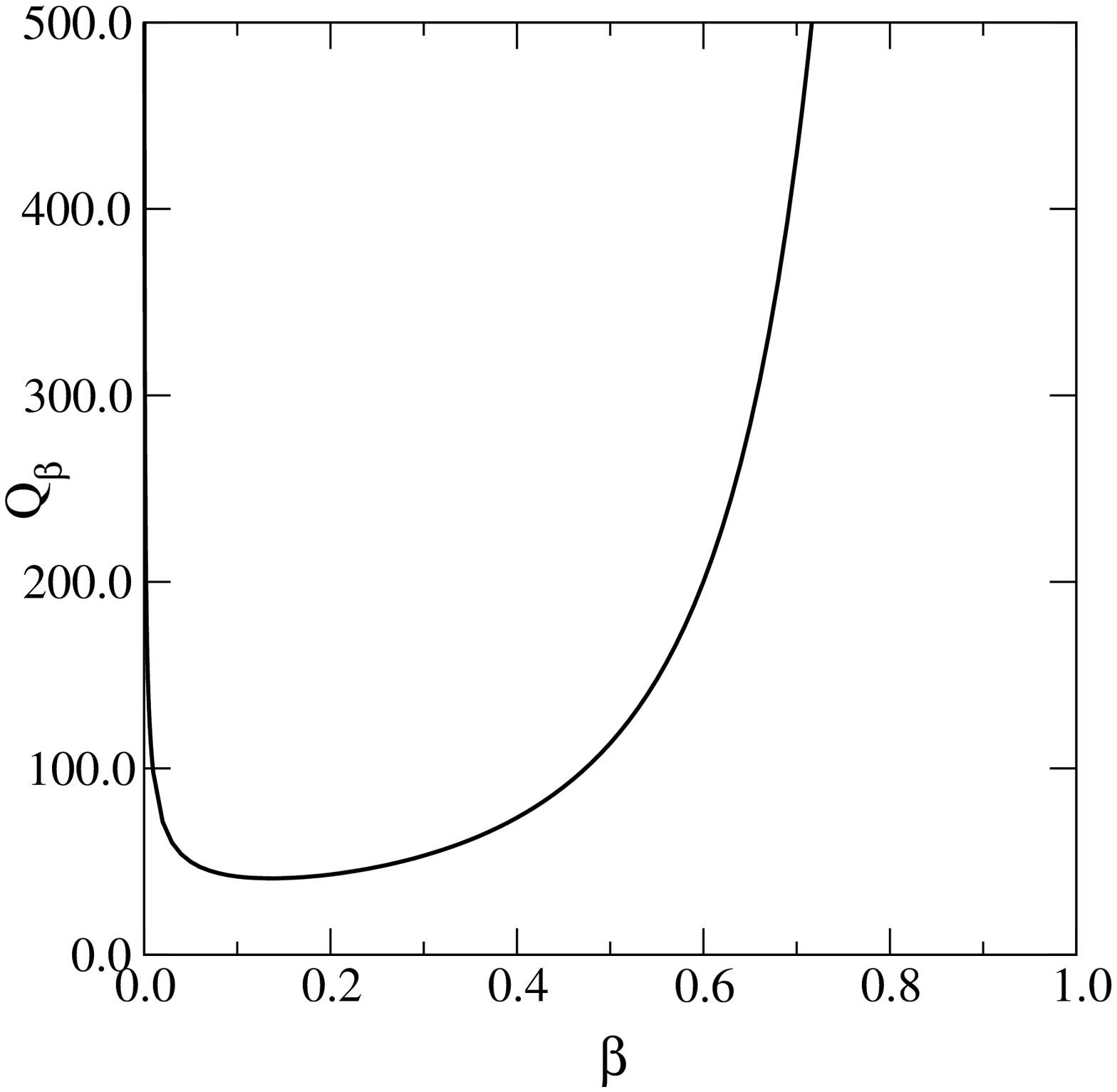,width=7cm}%
\hspace*{1cm}%
\epsfig{file=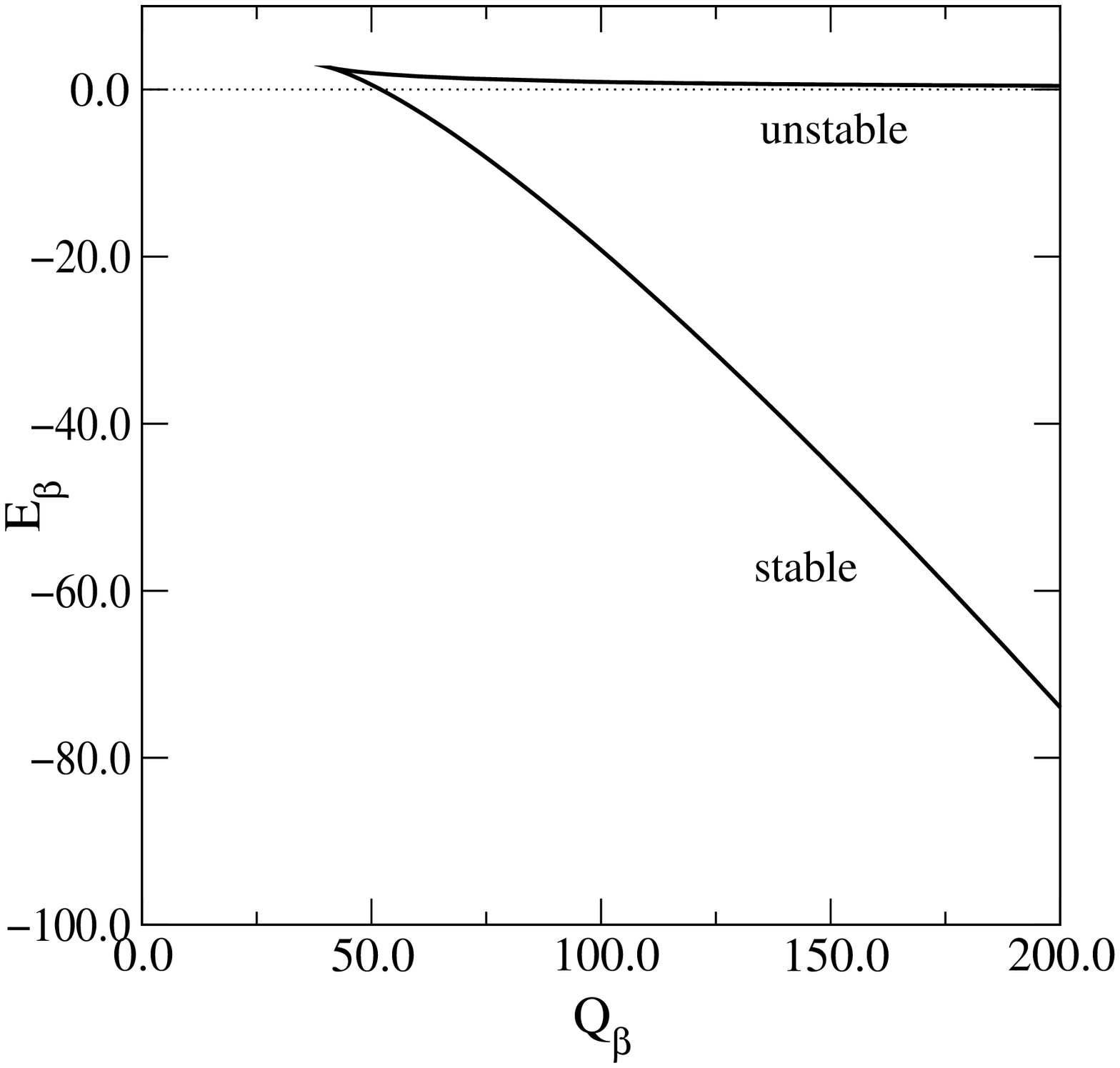,width=7cm}%
\end{center}

\caption[a]{Left: the rescaled charge, $Q_\beta$
(\eq\nr{Qbeta}), as a function
of $\beta$ (\eq\nr{rescale}). The two branches of solutions,
obtained with small and large $\beta$, are well
visible. Right: the rescaled energy of the Q-ball, $E_\beta$ (\eq\nr{Ebeta}),
as a function of $Q_\beta$ (\eq\nr{Qbeta}). The upper branch
(corresponding to $\beta < 0.14$) leads to unstable solitons,
while the lower branch ($\beta > 0.14$) leads to
metastable ($E_\beta > 0$) or stable ($E_\beta < 0$) ones.}

\la{fig:branches}
\end{figure}

\subsection{Analytic considerations}
\la{se:a}

Let us next discuss
some analytic estimates in the regimes of small and large $\beta$.
The regime of small $\beta$, or small $B$, is called the ``thick-wall''
regime: there the Q-ball resembles a lump without a separate interior
and a boundary; in other words, the boundary (or ``wall'') is as thick
as the radius. Some properties of the solution, such as its behaviour
at large $\hat r$, have previously been discussed, e.g., in~\cite{lp,mv}.
The quantity $Q_\beta$ behaves as $Q_\beta = \hat Q/\beta^{1/2}$,
where $\hat Q \approx 9.4486$. From \eqs\nr{rescale},
\nr{rescale2}, \nr{rsQmu},
this gives the relation
\be
 Q(|\mu|) =
 \Bigl( \frac{\hbar^2}{2 m |\mu|} \Bigr)^{1/2}
 \frac{\hat Q}{4 \pi |a|}
 \;. \la{Qthick}
\ee
Inserting into~\eq\nr{rescale}, the physical distance
is obtained from $\hat r$ as
\be
 r  = {4\pi |a|} \hat r \cdot
 \frac{Q}{\hat Q} 
 \;,
 \la{rell}
\ee
and the particle number density at a given distance is given
by
\be
 n(r)  =
 |\psi|^2 (r) = \fr12 {\hat v}^2(r) \cdot
 \biggl( \frac{\hat Q}{Q} \biggr)^2
 \biggl( \frac{1}{4\pi |a|} \biggr)^3
 \;.
 \la{nr}
\ee
It is observed that the central density is {\it smaller} for larger
particle numbers, but the size of the solution is larger.
The solution remains weakly interacting even in the center,
\be
 n(r) |a|^3 \ll 1
 \;,
 \la{wi}
\ee
for large values of $Q$. Note that all dependence on $B$ has cancelled
in \eqs\nr{Qthick}--\nr{wi}.

These thick-wall solutions of the equations of motion are not stable, however.
Using $Q = -\partial \Omega(\mu)/\partial\mu$ and \eq\nr{nrEQ}, 
one can derive $E(Q)$ from the $Q(|\mu|)$ in~\eq\nr{Qthick}, to obtain
\be
 E(Q) = \Bigl( \frac{\hbar {\hat Q}}{4\pi |a|} \Bigr)^2
 \frac{1}{2 m Q} \;.
 \la{solEQ}
\ee
Note that in contrast to the relativistic case, this energy does not
contain the particle rest masses. The fact that the total
binding energy $E(Q)$
is positive, implies that this solution is an excited state.
In fact, it is not even metastable: following~\cite{fls}, one can
inspect $Q$-conserving field
variations around the solution ${\hat v}(\hat r)$,
and find that there is a direction in the field
space where even a small variation leads to a decrease of $E(Q)$.
Therefore, the thick-wall solutions correspond to an unstable branch,
as shown in~\fig\ref{fig:branches}. There is, however, another solution 
with the same charge but a lower energy, to which we now turn.

The so-called thin-wall regime is obtained as
$\beta = |\mu| B (m/(\pi\hbar^2|a|))^2$ approaches unity.
At the same time, $\hat v(0)$ approaches 2.0. In this limit the
core of the soliton is essentially in a homogeneous phase, corresponding
to the global minimum of the theory, and has a well-defined
boundary, or ``wall'', which is thin compared with the radius.
Following~\cite{lp,ss}, this solution could also be called a droplet
of Bose liquid. In this limit, clearly, physics depends in an essential way
on the value of the stabilising terms, 
in our case the coefficient $B$.

The properties of the thin-wall solution can again
be found in the standard way.
One may first compute the interface tension of a planar wall at $\beta = 1$,
$\hat \sigma = \int_0^2 {\rm d}\hat v \sqrt{2 [-P(\hat v)]} = 2/3$. Then
$\Omega_\beta$ may be approximated as a sum of a surface term,
$4 \pi \hat R^2 \hat \sigma$, and a volume term, and
extremising this expression allows to solve for the radius $\hat R$.
Consequently, $Q_\beta$ and $E_\beta$ are easily obtained,
to leading order in $1-\beta$. We find the charge
\be
 Q_\beta \approx 4 \pi \Bigl(\frac{2}{3}\Bigr)^4 \frac{1}{(1-\beta)^3}
 \;,
\ee
and the radius
\be
 R \approx
 \Bigl( \frac{mB}{2} \Bigr)^{1/2}
 \frac{1}{\pi\hbar |a|}
 \Bigl(
 \frac{ 3 Q_\beta}{ 8 \pi }
 \Bigr)^{1/3}
 \;.
\ee
The binding energy becomes
\be
 E(Q) = - \biggl( \frac{\pi \hbar^2 |a|}{m}\biggr)^2
 \frac{Q}{B} + \mathcal{O}(Q^{2/3}) \;,
\ee
and, being negative for large $Q$, 
the solitons are absolutely stable. 
At the same time, the central density becomes
\be
 n |a|^3 \approx \frac{\pi\hbar^2|a|^4}{m B}\;,
 \la{dens}
\ee
independent of $Q$. Thus the interactions
are no longer weak, for a small $B$.

To summarise, soliton solutions exist independent of the value of $B$.
They come, however, in two branches, and the branch which remains there
in the limit $B\to 0$ (thick-wall, or small~$\beta$) 
corresponds to unstable Q-balls. Therefore stabilising terms, for instance 
of the form in~\eq\nr{higherops}, are essential for
the properties of stable atomic Q-balls, just as they are in the 
relativistic case.  
The stabilising terms tend to lead, however,
to strong interactions in the interior of the soliton, which may in fact
resemble a Bose liquid rather than a dilute gas.

Finally, let us mention that according to~\fig\ref{fig:branches},
the Q-ball solutions have a minimum charge, $Q_\beta\approx 41$,
corresponding in physical units to
$Q_\rmi{min} \approx 41 (mB/(2\hbar^2))^{1/2}(2\pi|a|)^{-2}$.
For a vanishing stabilising term, therefore, $Q_\rmi{min} \to 0$, 
and the Q-ball could have any charge. Note, however, that 
quantum corrections become important for small $Q$~\cite{g}, 
and our classical analysis is no longer trustworthy. On the side
of large $Q$, it has been suggested~\cite{solito}
that there can also exist
a maximal charge, $Q_\rmi{max}$, beyond which the system undergoes
a phase transition to the stable (``Bose liquid'') phase. Whether
this can happen depends on the ensemble: in our case (large volume, 
fixed $Q$) it cannot, because the charge density in the stable phase
is so large (cf.\ \eq\nr{dens}) that there are simply not enough 
particles present to fill the whole volume with this phase: 
the Q-ball solution is in fact the optimal configuration, 
and absolutely stable.

%
\section{Solutions in harmonic traps, and experimental data}
\la{se:trap}

In experiments where atomic BECs are studied, space is not homogeneous,
but there is a harmonic
trap, characterised by the potential $V(\vec{x})$ in~\eq\nr{nrEOM}. This
modifies the solution in a qualitative way. We reiterate here the
situation for spherically (``3D'') and axially
(``1D'') symmetric potentials\footnote{%
  We follow standard terminology although it is not
  without the danger of some confusion:
  In the ``1D'' case the trap is {\em narrow} in {\em two}
  directions, while in the ``3D'' case it is narrow in three.}.
We do not present any new solutions but simply point out how
the Q-ball picture fits the 
qualitative pattern of the condensate behaviour 
in these traps.

\paragraph{3D case.}

Condensate solutions for a (nearly) spherical trap,
$V(\vec{x}) \equiv  m \omega_r^2 r^2/2$, were obtained
in~\cite{rhb}. Their essential properties are as follows. The trap
has a finite width, characterised by $l_r = \sqrt{\hbar/(m \omega_r)}$,
where $\omega_r$ could typically be in the range of 100~Hz or so.
Because of a finite $l_r$, the radius of the soliton cannot grow
freely, but is restricted. Therefore, as more particles condense,
the only way to accommodate them is to increase the
central density. This behaviour is opposite to either of the 
branches discussed in
\se\ref{se:a}. Therefore the trapped ``soliton'' belongs to
yet a different ``branch'' of solutions than
the genuine Q-balls. Another way to express the issue
is that in the genuine Q-ball solutions $\mu < 0$ (cf.\ \se\ref{se:homo}), 
while in the trap solution $\mu > 0$~\cite{rhb}.

Because of the growth of the central density, trapped solutions with
large charges are unstable. Indeed, once the central density increases
beyond a certain limit, various losses become
overwhelming~\cite{dew}, and the condensate collapses,
as is also observed experimentally~\cite{bsh}. After the collapse,
the condensate may start to grow again, only to experience yet
another collapse later on~\cite{collapse}. The collapse happens
when $Q\sim \mathcal{O}(l_r/|a|)$, imposing an upper limit on
the charge, or particle number, in the condensate.

\paragraph{1D case.}

In the 1D case the trap has a small
finite width only in two directions, but is
very long in one direction. It turns out that in this case genuine
solitons {\it can} be observed. The chemical potential corresponding
to them is negative, as in~\se\ref{se:homo}. This solution,
called a ``bright soliton'',
was discussed in detail already in~\cite{brite}\footnote{%
  In 1D solitons exist even for $a > 0$~\cite{dark}. These
  are sometimes called ``dark solitons'', and consist essentially of
  a kink solution, where the field is zero inside an otherwise
  homogeneous condensate.}.

The essential properties of the 1D solitons can easily
be deduced from the Q-ball results in~\se\ref{se:homo},
in the thick-wall limit $B\to 0$ (see also
\cite{ccr,pmh} and references therein). Let us now denote
by $l_r$ the transverse width of the trap, and by
$a_r, a_v$ the scaling factors
in~\eq\nr{rescale}.
In the expression for $Q$, then,
the homogeneous 3D relation $Q \sim \hat Q\, a_r^3 a_v^2$
gets replaced with $Q \sim \hat Q\, a_r l_r^2  a_v^2$.
Therefore, $Q \propto |\mu|^{1/2}$. Consequently, the radius now scales
as $r \propto a_r \propto |\mu|^{-1/2} \propto 1/Q$, the central
density as $v^2 \propto a_v^2 \propto |\mu| \propto Q^2$,
the grand canonical potential
as $\Omega(\mu) \propto a_r l_r^2 a_v^2 |\mu|\propto |\mu|^{3/2}$,
and the energy as $E(Q) = \Omega(\mu) - \mu \partial \Omega(\mu)/\partial\mu
\propto -\fr12 |\mu|^{3/2} \propto -Q^3$.
Because the binding energy is negative, these Q-balls are
absolutely stable compared with the gaseous phase, 
even for $B=0$. The central
density grows with particle number, however, which may still
cause an instability related to the practical experimental 
setup for large particle numbers,
like for trapped 3D solitons.

Let us now note that
once they have formed~\cite{bright,train},
it is possible to study experimentally the {\it collisions}
of such 1D solitons~\cite{train}. It is very interesting that
the collision results are qualitatively similar to what has been
found in numerical simulations of the relativistic case~\cite{collisions},
{\it viz.}, that Q-balls with opposite phases
repel each other~\cite{train,kha}. Such similarities may provide 
an exciting opportunity
for studying supersymmetry-based post-inflationary cosmology in the laboratory.

%
\section{Conclusions}

We have emphasised in this paper that a rigorous formal analogy exists
between the non-topological solitons, or Q-balls, of relativistic
field theories, and three-dimensional solitons that could be found in
atomic BECs with a negative s-wave scattering length.

In order to be {\it stable}, the three-dimensional BEC soliton
requires an additional stabilising term, beyond the usual
four-point interaction. The precise form of the stabilising term 
is not essential, however: as an example, we have considered three-body
scattering~\cite{bhh}, the strength of which we denoted by $B$. 
Whether a stabilising interaction of precisely this type could be obtained
in atomic BECs with a negative scattering length, either directly or
effectively as a consequence of some coupling of the atoms to external
fields, remains at present an open issue. If it exists, as 
could be argued from
general principles following~\cite{bhh}, 
or if some other stabilising mechanism 
takes over~\cite{ss,htcs} (ultimately even the hard core atomic repulsion 
should be sufficient), then the Q-ball is stable against decay 
into its quanta, the free atoms.

Without any stabilising term, the stationary three-dimensional soliton
still exists, but it has a finite lifetime. The lifetime is currently 
unknown, but it could be determined by solving the time-dependent
Gross--Pitaevskii equation around the solution we have presented here.

These considerations have implications on both contexts in which
Q-balls may appear. In the BEC case, it is an interesting question
whether the spherical Q-ball solutions in a homogeneous space, which
are quite different from the traditional 3D trapped ones discussed
in~\se\ref{se:trap}, could also be observed experimentally. This is
no doubt a challenging task.  In principle one could attempt to tune
the trap frequency to as small a value as possible while $a>0$, and
then tune $a$ negative, to collapse an almost homogeneous BEC into a
genuine Q-ball soliton.  Alternatively one could start with a
significantly prolongated trap holding a genuine 1D soliton, and then
slowly decrease the trap frequency in the transverse directions, to
try and restore spherical symmetry.

On the side of cosmology, where most of the interest in Q-balls has
been in recent years, the current understanding of their dynamics is
based on solving classical equations of motion. In the actual BEC
experiments, of course, the system contains also quantum and thermal
fluctuations, modifying the dynamics. Thus experimental results from
BECs can to some extent test which features of the dynamics are robust
with respect to these fluctuations. The existing 1D
experiments~\cite{train,kha} are very encouraging in this respect, but
it would of course be even more remarkable if they could be extended
to the (almost) homogeneous 3D case, as outlined above.

%
\section*{Acknowledgements}

K.E. thanks E.~Lundh for discussions on the properties of atomic BECs
and for drawing our attention to relevant BEC literature, and
M.~Mackie and K.-A.~Suominen for discussions. This work was partly
supported by the RTN network {\em Supersymmetry and the Early Universe},
EU contract no.\ HPRN-CT-2000-00152.


\end{document}